\documentclass[ALICE,manyauthors]{cernphprep}

\usepackage[comma,square,numbers,sort&compress]{natbib}
\usepackage{hyperref}
\usepackage{lineno}
\usepackage{xcolor}
\usepackage{textcomp}
\usepackage[T1]{fontenc}
\usepackage{xspace}
\usepackage{orcidlink}

\usepackage{xspace}

\newcommand{\PbPb}         {\mbox{Pb--Pb}\xspace}

\newcommand{\snn}          {\ensuremath{\sqrt{s_{\mathrm{NN}}}}\xspace}
\newcommand{\pt}           {\ensuremath{p_{\rm T}}\xspace}

\newcommand{\nineH}        {$\sqrt{s}~=~0.9$~Te\kern-.1emV\xspace}
\newcommand{\seven}        {$\sqrt{s}~=~7$~Te\kern-.1emV\xspace}
\newcommand{\twoH}         {$\sqrt{s}~=~0.2$~Te\kern-.1emV\xspace}
\newcommand{\twosevensix}  {$\sqrt{s}~=~2.76$~Te\kern-.1emV\xspace}
\newcommand{\five}         {$\sqrt{s}~=~5.02$~Te\kern-.1emV\xspace}
\newcommand{\twosevensixnn}{$\sqrt{s_{\mathrm{NN}}}~=~2.76$~Te\kern-.1emV\xspace}
\newcommand{\fivenn}       {$\sqrt{s_{\mathrm{NN}}}~=~5.02$~Te\kern-.1emV\xspace}

\newcommand{\GeVc}         {Ge\kern-.1emV/$c$\xspace}
\newcommand{\MeVc}         {Me\kern-.1emV/$c$\xspace}
\newcommand{\TeV}          {Te\kern-.1emV\xspace}
\newcommand{\GeV}          {Ge\kern-.1emV\xspace}
\newcommand{\MeV}          {Me\kern-.1emV\xspace}
\newcommand{\GeVmass}      {Ge\kern-.2emV/$c^2$\xspace}
\newcommand{\MeVmass}      {Me\kern-.2emV/$c^2$\xspace}

\newcommand{\Hee}          {\ensuremath{^{3}\mathrm{He}}\xspace}
\newcommand{\antiHee}
{\ensuremath{^{3}\overline{\mathrm{He}}}\xspace}
\newcommand{\HH}          {\ensuremath{\mathrm{t}}\xspace}
\newcommand{\antiHH}      {\ensuremath{\overline{\mathrm{t}}}\xspace}

\newcommand{\dd}     {\ensuremath{\mathrm{d}}\xspace}
\newcommand{\gmom}   {\mbox{\ensuremath{\mathrm{GeV}/c}}\xspace}

\begin{document}%

\begin{titlepage}
\PHyear{2022}
\PHnumber{263}      
\PHdate{21 November}  
%

\title{Light (anti)nuclei production in Pb--Pb collisions at $\sqrt{s_{\mathrm{NN}}}=5.02$ TeV}
\ShortTitle{Light (anti)nuclei production in Pb--Pb collisions at $\sqrt{s_{\mathrm{NN}}}=5.02$ TeV}   
\hyphenation{nu-cleo-syn-the-sis}
\Collaboration{ALICE Collaboration\thanks{See Appendix~\ref{app:collab} for the list of collaboration members}}
\ShortAuthor{ALICE Collaboration} 

\begin{abstract}
The measurement of the production of deuterons, tritons and \Hee and their antiparticles in \PbPb collisions at $\snn = 5.02$ TeV is presented in this article. The measurements are carried out at midrapidity ($|y| < $ 0.5) as a function of collision centrality using the ALICE detector.
The \pt-integrated yields, the coalescence parameters and the ratios to protons and antiprotons are reported and compared with nucleosynthesis models.
The comparison of these results in different collision systems at different  centre-of-mass collision energies reveals a suppression of nucleus production in small systems. In the Statistical Hadronisation Model framework, this can be explained by a small correlation volume where the baryon number is conserved, as already shown in previous fluctuation analyses. However, a different size of the correlation volume is required to describe the proton yields in the same data sets. The coalescence model can describe this suppression by the fact that the wave functions of the nuclei are large and the fireball size starts to become comparable and even much smaller than the actual nucleus at low multiplicities.
\end{abstract}
\end{titlepage}
\setcounter{page}{2}

\section{Introduction}
\label{sec:intro}
Collisions of ultrarelativistic heavy ions create suitable conditions for the production of light (anti)nuclei, as a high energy density is reached over a large volume. Under these conditions, hot and dense matter, which contains approximately equal numbers of quarks and antiquarks at midrapidity, is produced for a short duration (a few 10$^{-23}$\,s). After a deconfined quark--gluon plasma (QGP) is formed in the initial state, the system cools down and undergoes a transition to a hadron gas. While the hadronic yields are fixed at the moment when the rate of inelastic collisions becomes negligible (chemical freeze-out), the transverse momentum (\pt) distributions continue to change until elastic interactions cease (kinetic freeze-out). The observed nucleus abundance is highly sensitive to the chemical freeze-out conditions as well as the dynamics of the emitting source. 

The production of light nuclei and antinuclei has already been measured in many experiments at various energies in heavy-ion collisions at the Bevalac~\cite{Bevalac81}, SIS~\cite{FOPI:2010xrt,HADES:2020lob},  AGS~\cite{E886,E864,E864-dbar,E896,E878:1998vna},  SPS~\cite{NA44,NA49-d,NA49,NA52:2003udo}, RHIC~\cite{Adler:2001uy,PHENIX2005,Afanasiev:2007tv,Abelev:2009ae,Agakishiev:2011ib,Adam:2019wnb,STAR:2020hya,STAR:2021ozh} and LHC~\cite{ALICE:2015rey,Adam:2015vda,ALICE:2017nuf,ALICE:2022zuz,ALICE:2022amd} and in smaller collision systems~\cite{ISR,ISR-2,Alexopoulos2000,H1,CLEO-2007,ALEPH,Adam:2015vda,ALICE:2017xrp,ALICE:2019bnp,ALICE:2019dgz,ALICE:2020foi,ALICE:2020hjy,ALICE:2021mfm,ALICE:2021ovi}. The production of light nuclei is usually discussed within two theoretical approaches: the nucleon coalescence model~\cite{Kapusta:1980zz,Mrowczynski:1987oid,Mrowczynski:1989jd,Scheibl:1998tk,Steinheimer:2012tb,Bellini:2018epz} and the statistical hadronisation model (SHM)~\cite{BraunMunzinger:2003zd,Andronic:2010qu,Cleymans:2011pe,Andronic:2017pug,Donigus:2020ctf}.  

Generally, the coalescence model describes the production of nuclei from the nucleons emitted from a hot fireball that cools down while expanding. Most variants use a phase-space picture for the formation, namely the nucleons have to be close in space and (relative) momentum to allow for the formation of the nucleus.
In particular, the studies of the production of (anti)nuclei as a function of the charged-particle multiplicity~\cite{Adam:2015vda} have clearly shown experimentally a dependence of the yield (ratios) on the volume of the emitting fireball. In fact, a strong suppression of nucleus-to-proton yield ratios is seen from high multiplicities in Pb--Pb to lower multiplicities in p--Pb, and reaching small multiplicities in pp~\cite{ALICE:2021ovi,ALICE:2021mfm} collisions. The size of the fireball can be extracted from the measurement of two-particle correlations, of Hanbury Brown \& Twiss type~\cite{hbt56a,hbt56b,GoldhaberHBT}, nowadays often called femtoscopy. Empirically, it was found that the charged-particle multiplicity of the collision is proportional to the size parameter $R$ cubed~\cite{ALICE:2015hav,Bellini:2018epz}.  
Theoretically, these correlations can be directly connected to the coalescence parameter $B_A$, which is a measure for the probability to form a nucleus of mass number $A$ from the corresponding nucleons~\cite{Mrowczynski:1987oid,Mrowczynski:1989jd,Nagle:1996vp,Scheibl:1998tk,Blum:2017iwq,Blum:2017qnn,Bellini:2018epz}.   
This dependence is mainly given by the fact that the wave functions of the nucleons have to overlap with the nucleus' wave function, while the constituents are being emitted from a region of homogeneity of the fireball. 
This leads to a strong suppression of nucleus production in small systems, since the size of the formed nucleus becomes larger than the emitting source~\cite{Sun:2018mqq,Bellini:2018epz}. 
It is worthwhile to mention that coalescence models typically ignore the problem of conservation of energy and momentum in the process, assuming either that this is handled via the off-shell nature of the particles or additional particles involved in the process itself.

The SHM successfully describes hadron yields in heavy-ion collisions, in particular in central collisions and at midrapidity~\cite{Andronic:2017pug}. 
The production of nuclei is solely determined by their quantum numbers and masses~\cite{Donigus:2020ctf}. For more peripheral heavy-ion collisions or even smaller collision systems such as pp and p--Pb, one needs to switch from a grand canonical ensemble to a canonical description of the relevant quantum numbers (baryon number $B$, charge $Q$, and strangeness $S$)~\cite{Vovchenko:2018fiy,Vovchenko:2019kes}. The canonical ensemble requires a local conservation of each quantum number in a particular volume $V_\mathrm{c}$, the so called correlation volume. Interestingly, $V_\mathrm{c}$ cannot be unambiguously determined from first principles~\cite{Castorina:2013mba}, but it can be constrained from measurements of the event-by-event number fluctuation of net protons~\cite{Vovchenko:2019kes,ALICE:2019nbs} or  deuterons~\cite{ALICE:2022amd}.
Other approaches include a non-equilibrium treatment of the quantum numbers in question (see Ref.~\cite{Donigus:2020ctf} and references therein) or even a partial chemical equilibrium (PCE)~\cite{Xu:2018jff,Vovchenko:2019aoz,Neidig:2021bal}.

In this article, the production of deuterons, tritons and \Hee and their antiparticles in \PbPb collisions at a centre-of-mass energy per nucleon pair $\snn = 5.02$ TeV is reported. The \pt-integrated yields, the coalescence parameters (calculated using average of protons and antiprotons, since they are found to be produced in same abundance at the LHC energies~\cite{ALICE:2019hno}.) and the ratios to protons and antiprotons are compared with nucleon coalescence and statistical hadronisation models.

\section{Experimental apparatus and data sample}
\label{sec:detector}
The results presented in this article are based on the data set of \PbPb collisions at $\mathbf{\sqrt{{\textit s}_{\rm NN}}} = 5.02$ TeV collected in 2018.
In total, $230\times10^6$ events were analysed, of which $86.7\times10^6$ are central trigger events in the 0--10\% centrality interval and $74.3\times10^6$ are semicentral trigger events in the 30--50\% centrality interval.

The ALICE detector~\cite{Aamodt:2008zz,Abelev:2014ffa} has excellent particle identification and vertexing capabilities.
The \mbox{(anti)}~nuclei were measured using the Inner Tracking System (ITS), the Time Projection Chamber (TPC) and the Time-Of-Flight (TOF) detector.
All these detectors are located inside a homogeneous magnetic field with a strength of 0.5~T and cover the full azimuthal acceptance and the pseudorapidity range $|\eta|<0.9$ for interactions located in $|z|<$ 10 cm, where $z$ is the distance from the nominal interaction point along the beam direction.

The ITS~\cite{ITS2010} consists of six cylindrical layers of (position-sensitive) silicon detectors, covering the central rapidity region.
The ITS allows the reconstruction of the primary and secondary
vertices. It is also used to separate primary nuclei from secondary nuclei via the distance of closest approach (DCA) of the track to the primary vertex with good resolution (better than 300 $\mu$m), assured by the Silicon Pixel Detector (SPD), which comprises the innermost two layers of the ITS.

The TPC~\cite{Alme2010} is the main tracking device of the experiment. It is a gas-filled cylinder and provides particle identification via the specific energy loss (d$\textit E$/d$\textit x$). (Anti)$^3$He are identified up to $\pt = 7~\gmom$ using the TPC only.

The TOF detector~\cite{TOF2013} allows for the light (anti)nuclei identification by means of the velocity determination.
Its total time resolution for tracks from \PbPb collisions corresponds to about 65 ps which is determined by the intrinsic time resolution of the detector and the resolution of the event collision time measurement.
By a combined analysis of TPC and TOF data, (anti)deuterons are identified up to $\pt = 6~\gmom$ in \PbPb collisions. (Anti)tritons are also identified using TPC and TOF. However, due to a sizeable background starting at about 2~\gmom originating from mismatches between a track and a cluster in TOF, the (anti)tritons can only be measured up to $\pt = 3.2~\gmom$ in this data set.

The Transition Radiation Detector (TRD)~\cite{ALICE:2017ymw} was designed to provide electron identification and triggering and to improve the track reconstruction and calibration in the central barrel of ALICE. 
The TRD improves the overall momentum resolution of the ALICE central barrel by providing additional space points at large radii for tracking, reducing as well significantly the probability of mismatch between tracks and TOF hits for rare probes analyses such as the triton analysis presented in this article.

Finally, a pair of forward and backward scintillator hodoscopes ($2.8<\eta<5.1$ and $-3.7<\eta<-1.7$), the V0 detectors~\cite{VZERO2013},
measures the arrival time of particles with a resolution of 1 ns. The V0 detectors are used for triggering purposes and for rejection of beam–-gas interactions.
Furthermore, it provides the centrality determination in Pb--Pb collisions.

\section{Data analysis}
\label{sec:analysis}
\subsection{Event and track selection}
\label{evtrkSel}
The data were collected using a minimum-bias trigger requiring at least one hit in both the V0 detectors. In addition, a central and a semicentral trigger were used, also determined by the V0 detectors,  selecting collisions in the 0--10\% and 30--50\% centrality intervals, respectively.
Moreover, the timing information of the V0 scintillator arrays is used to reject the events triggered by the interactions of the beam with the residual gas in the LHC vacuum pipe. A further selection using the Zero Degree Calorimeter is applied in order to reject the electromagnetic beam--beam interactions and beam--satellite bunch collisions~\cite{Herr:941319}. These three rejections are done in the offline analysis.

The production yield of primary (anti)deuterons, (anti)tritons and (anti)$^3$He are measured at midrapidity. In order to provide optimal particle identification by reducing the difference between transverse and total momentum, the spectra are provided within a rapidity window of $|y|\,<$ 0.5.
Only tracks in the full tracking acceptance of $|\eta|\,<$ 0.8 are selected. 
In order to guarantee good track momentum and d$\textit E$/d$\textit x$ resolution in the relevant $p_{\mathrm{T}}$ ranges, the selected tracks are required to have at least 70 out of 159 possible reconstructed points in the TPC and two points in the ITS (out of which at least one is in the SPD). 
The requirement of at least one point in the two innermost layers, the SPD, assures a resolution better than 300 $\mu$m  on the distance of closest approach to the primary vertex in the plane perpendicular (DCA$_{xy}$) and parallel (DCA$_z$) to the beam axis for the selected tracks~\cite{Abelev:2014ffa}.
Furthermore, it is required that the $\chi^{2}$ per TPC reconstructed point is less than 2.5 and tracks of weak-decay products are rejected as they cannot originate from the tracks of primary nuclei.

\subsection{Particle identification}
\label{partId}

\begin{figure}[tp]
  \centering
  \includegraphics[width=0.7\textwidth]{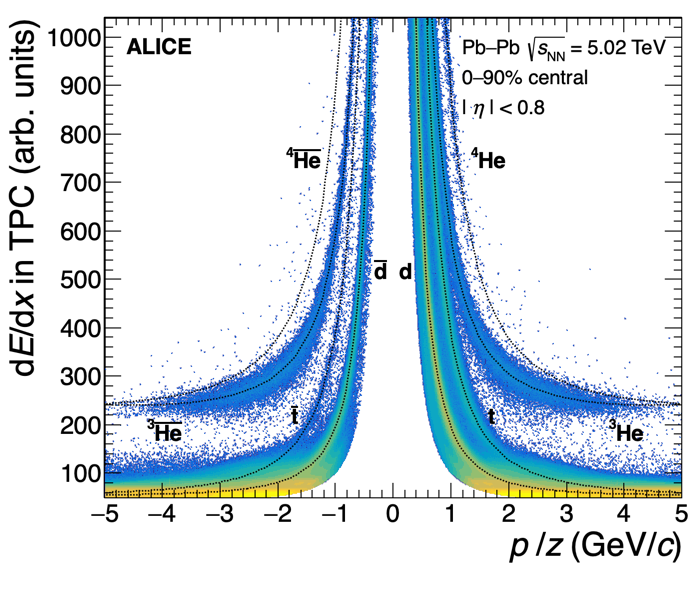}
  \caption{Specific energy loss of charged tracks in the TPC vs.\,rigidity ($p/z$) for Pb--Pb collisions at $\mathbf{\sqrt{{\textit s}_{\rm NN}}} = 5.02$ TeV. The dashed lines represent parameterisations of the Bethe--Bloch curve. Particles lighter than deuterons have been removed by applying a selection in the d$E$/d$x$ vs. $p/z$ plane that corresponds to the upper edge (3$\sigma$) of the proton band such that only nuclei are visible.}
  \label{fig:tpcdedx}
\end{figure}

The TPC allows for a clean identification of (anti)$^3$He in the whole \pt range and of (anti)deuterons up to \pt$\approx 1$ GeV/$c$. For higher transverse momenta, the d$E$/d$x$ information for charged particles is combined with the TOF mass determination in the (anti)deuteron analysis. For the (anti)tritons in the whole \pt range, a combined TPC and TOF analysis is performed.
Figure~\ref{fig:tpcdedx} shows the TPC specific energy loss as a function of rigidity ($p/z$) in Pb--Pb collisions at $\mathbf{\sqrt{{\textit s}_{\rm NN}}} = 5.02$ TeV.
The dashed curves represent parameterisations of the Bethe--Bloch formula for the different particle species.
The (anti)deuteron and (anti)$^3$He identification with the TPC is achieved by requiring that the energy-loss signal of a track lies in a 3$\sigma$ window around the expected value for a given mass hypothesis, where $\sigma$ is the d$E$/d$x$ resolution. For the (anti)tritons, a reduced 2$\sigma$ window is employed in order to further decrease the background. 

\begin{figure}[tp]
 \centering
 \includegraphics[width=0.49\textwidth]{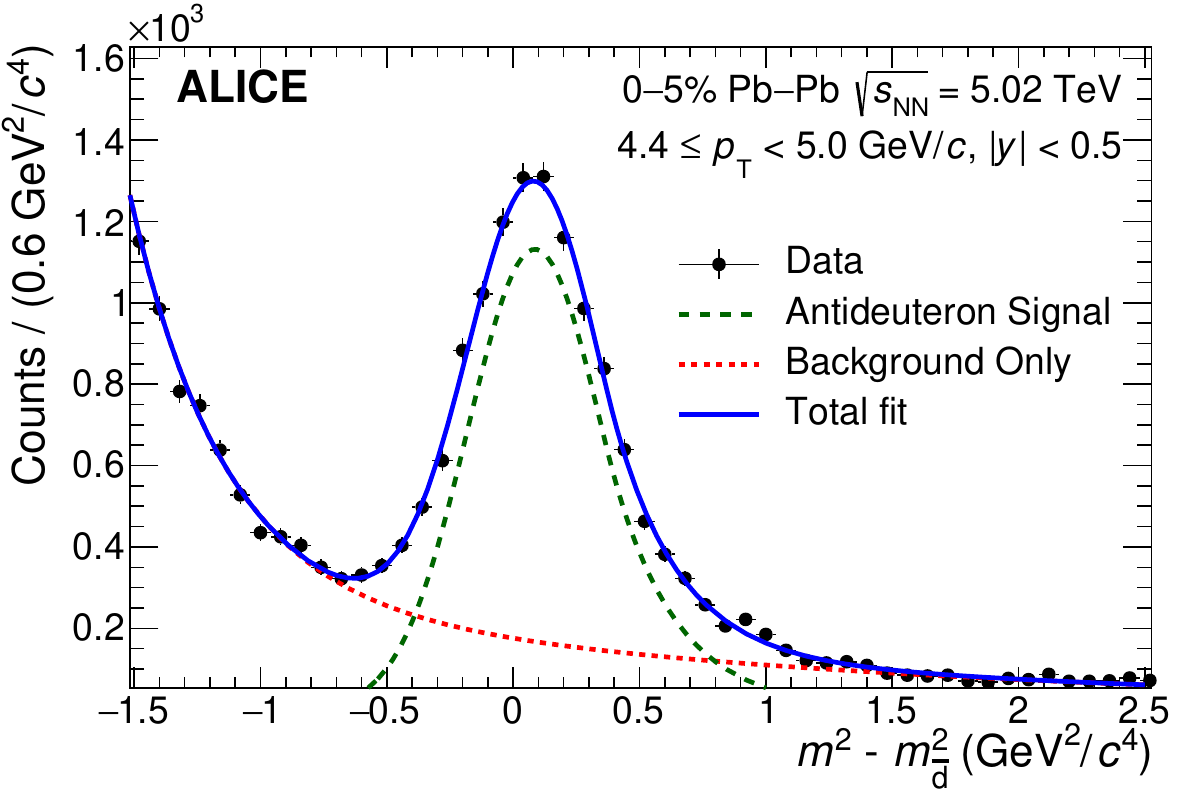}
 \includegraphics[width=0.49\textwidth]{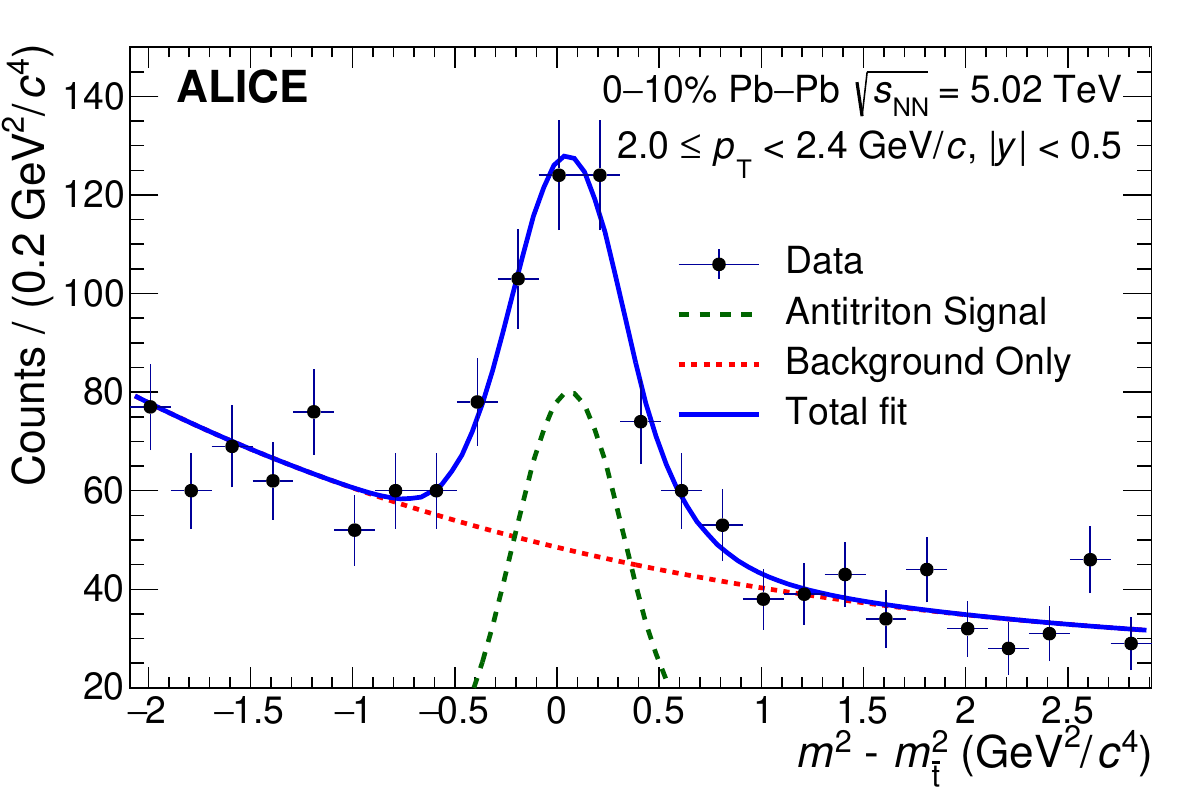}
 \caption{Fit to the measured squared mass to extract the antideuteron signal in 4.4 $<$ \pt $<$ 5.0 GeV/$c$ (left) and the antitriton signal in 2.0 $<$ \pt $<$ 2.4 GeV/$c$ (right). The red dashed line shows the background, the solid blue line the combined fit to the data and the green dashed line the signal only.}
 \label{fig:m2deuteron}
\end{figure}

In order to extend the \pt reach of the measurement, it is additionally required that the track is matched to a hit in the TOF detector.
As shown in Fig.~\ref{fig:m2deuteron}, based on the time-of-flight measurement the squared mass of the particle is determined in different \pt intervals and 
the distributions are then fitted using a Gaussian function with an exponential tail for the signal.  
The background of the (anti)deuterons mainly originates from two components, namely wrong association of a track with a TOF hit and the non-Gaussian tail of lower mass particles. For the (anti)tritons the dominant background originates from the wrong associations of a track with a TOF hit. 
For both nuclei, the background is described with the sum of two exponential functions.

\subsection{Background rejection}
\label{bckrej}

One of the main sources of background in the analyses of the primary deuteron and triton production are nuclei originating from secondary interactions.
These secondary nuclei come mostly from the interactions of other primary particles with the detector material.
In some of these interactions, a light nucleus can be produced by spallation processes, i.e. can be knocked-out from detector or from support material.
The baryon number conservation sets a very high energy threshold for the production of secondary antinuclei with similar processes, thus making the contribution of secondary antinuclei from material negligible, as also confirmed by simulations.
Other processes, such as the decay of (anti)hypernuclei, represent a negligible contamination to the observed (anti)deuterons and (anti)tritons.

\begin{figure}
	\begin{center}
		\includegraphics[width=0.95\textwidth]{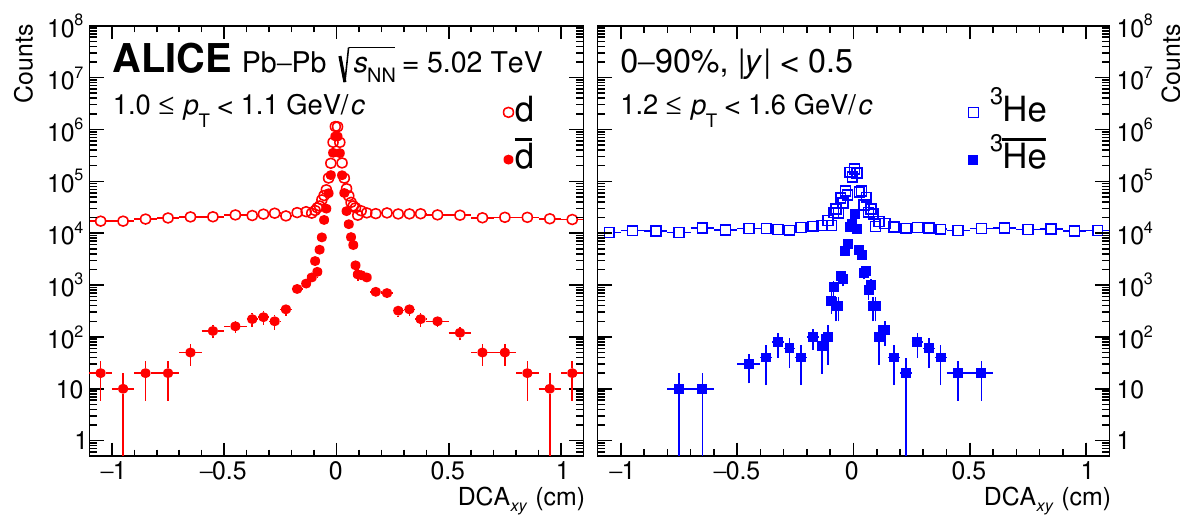}
		\caption{\label{fig:fit-dcaxy}$\rm{DCA}_{\it{xy}}$ of deuterons (open red circles) and antideuterons (solid red circles) for the $\pt$ intervals $1.0 \leq \pt < 1.1~\rm{GeV}/\it{c}$ (left) and of \Hee (open blue squares) and \antiHee (solid blue squares) for $1.2 \leq \pt < 1.6$ $\rm{GeV}/\it{c}$ (right).}
\end{center}
\end{figure}

As already done in previous analyses~\cite{ALICE:2015rey,Adam:2015vda,ALICE:2017nuf,ALICE:2022zuz,ALICE:2022amd}, in order to subtract the background from secondary deuterons and \Hee the DCA$_{xy}$ is used.
The distribution of primary particles is expected to be peaked at DCA$_{xy}\,=\,0$, whereas secondary particles are expected to exhibit a flat DCA$_{xy}$ distribution to the first order. The typical distributions of DCA$_{xy}$ for nuclei and antinuclei detected in ALICE are shown in Fig.~\ref{fig:fit-dcaxy}.
In second order, the tracks originating from secondary particles may be associated to a wrong hit in the innermost layers of the ITS.
If the latter belongs to a primary particle, the extrapolation of the secondary particle track will wrongly point to the primary vertex, as the track pointing is mostly driven by the hits in the innermost layers of the ITS.
In the deuteron and \Hee analyses presented in this article, a fit to the observed DCA$_{xy}$ distribution is performed to extract the primary fraction of deuterons and \Hee.
The DCA$_{xy}$ distributions of primary and secondary deuterons as well as \Hee in each transverse momentum interval are extracted from Monte Carlo (MC) events and are used as templates to fit the measured DCA$_{xy}$ distribution.
Since the secondary particles have large DCA$_{xy}$, the fits are done in a range of DCA$_{xy}$ wider than the actual track selection criterion to better constrain the secondary particle components.
The contamination from deuterons produced in the interactions with the detector material is only significant below 1.4 \gmom. 

In contrast to deuterons, the background from secondary tritons is rather dominant over the low number of primary triton counts. As this background only occurs at low \pt, the triton yield is only measured above 2.4~\gmom (2.0~\gmom in the most peripheral centrality interval).

\subsection{Corrections to the spectra}
\label{corrections}

The $p_{\mathrm{T}}$-differential-production spectra of (anti)deuterons, (anti)\Hee and (anti)tritons are obtained by correcting the raw spectra for tracking efficiency and acceptance based on MC generated events.
The MC samples used to compute the efficiency and the acceptance corrections for the Pb--Pb analysis were generated using the HIJING event generator~\cite{Wang:1991hta}. 
Since HIJING does not provide light (anti)nuclei, an \emph{ad hoc} generator that injects particles on top of the event generator was used.
The kinematics of the injected nuclei is chosen randomly by picking their transverse momentum from a flat distribution in the range between 0 and 10 \gmom, their azimuthal angle from a flat distribution between 0 and 2$\pi$ radians, and their rapidity from a flat distribution in the range $|y|\,<\,1$.
All particles are transported with GEANT4~\cite{Agostinelli:2002hh} through a full simulation of the ALICE detector.
The GEANT4 version used in the ALICE software framework was modified to take into account the latest (anti)nuclei hadronic interaction measurements~\cite{ALICE:2020zhb,ALICE:2022zuz}.

For the (anti)deuteron, (anti)\Hee and (anti)triton analyses, the efficiency$\times$acceptance was determined for each centrality interval separately. 
The input \pt distributions of (anti)nuclei in the simulation are modified according to a Blast-Wave parametrization using \pt-dependent weights. The BW parameters are taken from~\cite{ALICE:2019hno}.

\section{Systematic uncertainties}
\label{sec:Syst}
The sources of systematic uncertainties affecting these measurements were studied as follows:
\begin{enumerate}
\item the amount of material budget employed in the MC simulation of the ALICE apparatus was varied by $\pm$ 4.5\%, corresponding to the uncertainty on the ALICE material budget determination~\cite{Abelev:2014ffa};
\item track selection criteria were varied as done for previous analyses~\cite{ALICE:2015rey,Adam:2015vda,ALICE:2017nuf,ALICE:2022zuz,ALICE:2022amd};
\item the fit functions used for the signal extraction were varied;
\item for the antitriton analysis different functions were used to weight the input spectra in the simulation.
\end{enumerate}

A large contribution of the systematic uncertainty is due to the limited knowledge of the interaction of nuclei with the detector material. The main transport code used in ALICE is GEANT4~\cite{Agostinelli:2002hh}. This is used to estimate the efficiency$\times$acceptance that is applied as correction to the spectra. 
In general, the accuracy of the transport codes is limited by the available data for nuclei and especially (anti)nuclei hadronic interaction cross sections, which have only been measured in energy ranges far from the typical momenta of light (anti)nuclei produced in heavy-ion collisions~\cite{Auce:1996fe,Binon:1970yu,Jaros:1977it,Denisov:1971im}.
In a detailed study comparing the available data on different targets and their description in GEANT4, the corresponding uncertainty is evaluated as the scaling factor for the cross section value in GEANT4 that is required to match the experimental data. The simulations to estimate the efficiency are then repeated with the cross section in GEANT4 scaled by this factor. The determined systematic uncertainty is below 1\% for the deuterons using TPC and TOF and about 8\% for antideuterons at low \pt, and decreases down to 4\% at the largest \pt.
For \Hee it is about 0.5\% and for \antiHee it is about 2\% with a small dependence on \pt. The values for \HH and \antiHH are similar, with about 0.5\% and between 2.5\% and 5\%, respectively. 
In addition, weak decays from (anti)hypertritons can affect the (anti)\Hee spectra and contribute to the systematic uncertainties with about 1.7\%.
The discrepancy between the data and MC description of the ITS--TPC matching efficiencies is accounted for by adding 5\% of systematic uncertainties. All the other systematic uncertainties in the Pb--Pb analyses were estimated separately for each centrality class: particle identification and analysis selection criteria contribute by less than 3\%; the signal extraction method by less than 2\%; the TPC particle identification systematic uncertainty is estimated to be less than 2\%.

The huge background and the low number of counts of the (anti)triton analysis result in quite large statistical uncertainties and the systematic variations were found to be not significant within the statistical uncertainties. Therefore, the uncertainties from the variations were dropped and instead systematic uncertainties based on similar studies for charged pions, kaons, and protons were assigned, namely 5\% for the ITS--TPC matching efficiency and 6\% for the signal extraction for all centrality and \pt intervals.
The uncertainty on the \pt spectra coming from the uncertainty of the ALICE material budget was determined to be 2\%~\cite{ALICE:2022zuz}.
For the weighting of the efficiency$\times$acceptance, various functional dependencies were applied. This results in a negligible uncertainty in the higher \pt intervals, where the weighting does not have any effect, and up to 8\% uncertainty in the lower \pt intervals.  
To evaluate the systematic uncertainty due to the background estimation, an alternative data-driven method was used to determine the background. In this method, non-triton candidates were selected in the TPC by requiring
that their d$E$/d$x$ is outside a $\pm 2 \sigma$ window around the triton peak. Their squared TOF-mass distribution is used as a template for the background which is scaled to the squared TOF-mass distribution of triton candidates. It matches very well the background outside the triton peak region and allows an independent estimate of the background under the triton peak. This results in no systematic uncertainty in the low \pt region without any background and up to 20\% in the higher \pt intervals. 
All these contributions result in a total systematic uncertainty for the (anti)triton \pt spectra between 8\% and 22\%.

\section{Results}
\label{sec:results}
\begin{figure}[!ht]
	\begin{center}
		\includegraphics[width=0.98\textwidth]{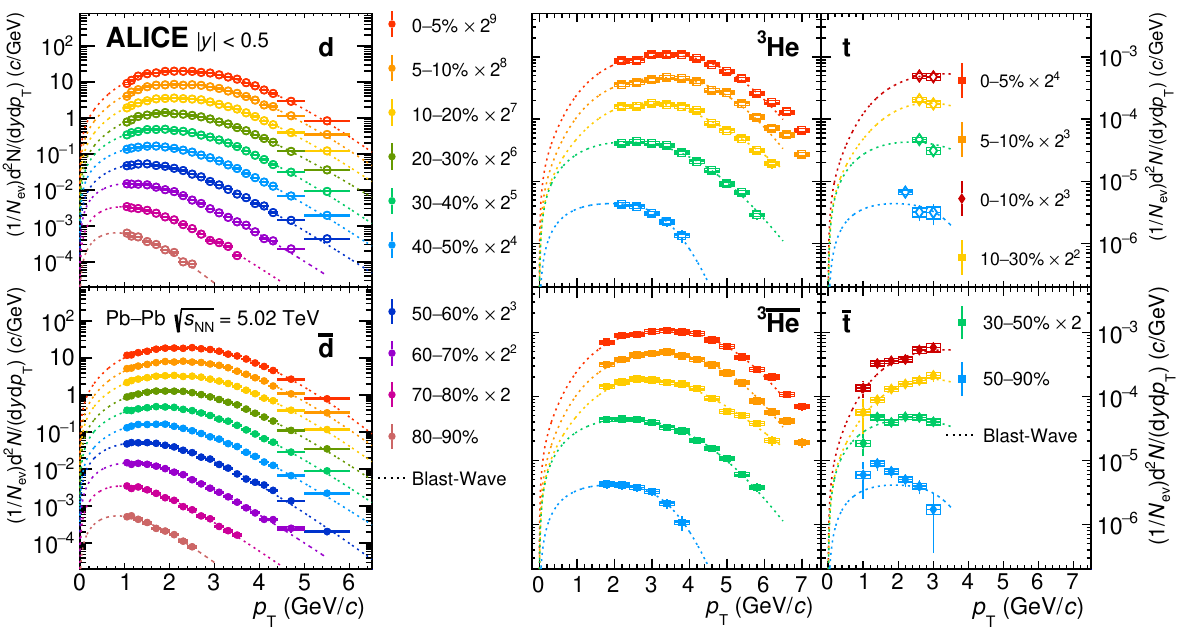}
		\caption{\label{fig-c5:spectra-d502}(Anti)deuteron, (anti)\Hee and (anti)\HH spectra measured in Pb--Pb collisions at $\mathbf{\sqrt{{\textit s}_{\rm NN}}} = 5.02$ TeV for different centrality classes reported with different colours.
		The boxes represent the systematic uncertainties, while the vertical lines are the statistical ones. The dashed lines represent the individual Blast-Wave fits to the spectra. The Blast-Wave fits of (anti)\Hee are used on (anti)\HH spectra as well to show the trend.}
	\end{center}
\end{figure}

The transverse momentum spectra of (anti)deuterons, (anti)\Hee and (anti)tritons are extracted in Pb--Pb collisions at $\mathbf{\sqrt{{\textit s}_{\rm NN}}} = 5.02$ TeV for various centrality classes. 
The transverse momentum spectra are shown in Fig.~\ref{fig-c5:spectra-d502}. A clear evolution of the spectral shape with centrality is observed, with the average transverse momentum almost doubling its value going from peripheral to most central \PbPb collisions and a shift in the peak position towards higher \pt for increasing multiplicity.

In order to measure the total yield per rapidity unit in Pb--Pb collisions, the spectra were fitted with the Blast-Wave function, which assumes a thermal production of particles from an expanding source~\cite{Schnedermann:1993ws}.
The systematic uncertainty of the integrated yield is obtained by shifting the spectrum within its systematic uncertainties and adding an additional uncertainty quadratically to account for the extrapolation to low and high $p_{\mathrm{T}}$. The latter is estimated by using different fit functions such as the $m_{\mathrm{T}}$ exponential, Boltzmann, Fermi--Dirac, and Bose--Einstein functions~\cite{stathandbook}. 
The fractions of extrapolated yield at low $p_{\mathrm{T}}$ for different centrality classes are about 5\% to 40\% for (anti)deuterons, 15\% (8\%) to 50\% (35\%) for (anti)\Hee and 23\% (1\%) to 50\% (11\%) for (anti)tritons depending on the centrality class. The extrapolated yields in the high-$p_{\mathrm{T}}$ region are negligible for most of the centrality classes except for a 3\% contribution for the most peripheral collisions for (anti)deuterons and (anti)\Hee, and a 55\% to 15\% contribution depending on the centrality class for (anti)tritons.
The statistical uncertainties are calculated by repeating the Blast-Wave fit by shifting the spectra randomly with a Gaussian distribution within the statistical uncertainties of each \pt interval. The resulting yield distribution is fitted with a Gaussian and the width of this distribution is taken as the statistical uncertainty.

\begin{figure}[!t]
	\begin{center}
		\includegraphics[width=\textwidth]{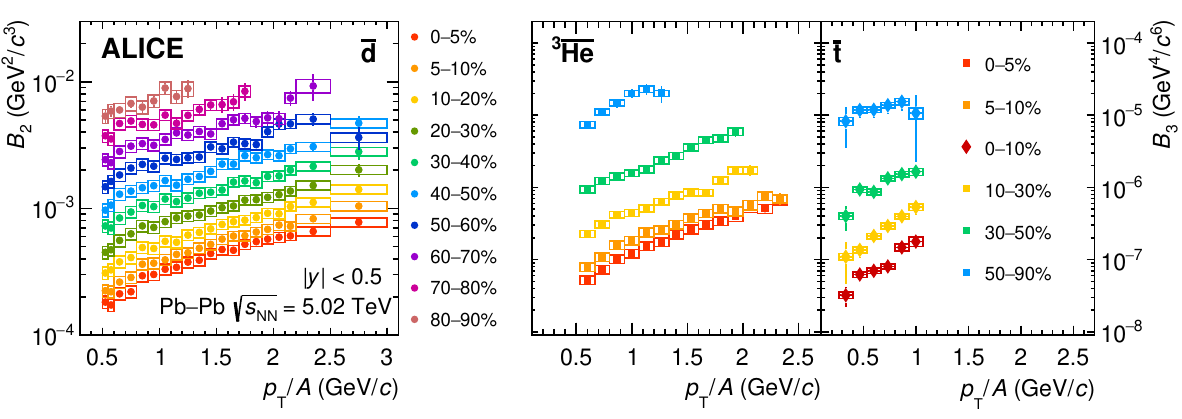}
		\caption{\label{fig-c5:coalescence}Coalescence parameters $B_2$ (left) and $B_3$ (right)  measured for antinuclei in Pb--Pb collisions at $\mathbf{\sqrt{{\textit s}_{\rm NN}}} = 5.02$ TeV as a function of the transverse momentum scaled by the mass number. Compatible results are obtained for the nuclei. Each colour corresponds to a different centrality class. The boxes represent the systematic uncertainties, while the vertical lines are the statistical ones. See the text for details.}
	\end{center}
\end{figure}

The coalescence scenario can be tested by computing the coalescence parameter $B_A$ (see for instance Ref.~\cite{Braun-Munzinger:2018hat} and references therein). Under the assumption of equal production of protons and neutrons, it is defined as
\begin{equation}
	B_A=E_A\frac{\dd^3N_A}{\dd p^3_A}\left(E_{\mathrm{p}}\frac{\dd^3N_{\mathrm{p}}}{\dd p^3_{\mathrm{p}}}\right)^{-A},
\end{equation}
where $E_A\frac{\dd^3N_A}{\dd p^3_A}$ and $E_{\mathrm{p}}\frac{\dd^3N_{\mathrm{p}}}{\dd p^3_{\mathrm{p}}}$ are the invariant production spectra of the nuclei with mass number $A$ and of protons, respectively. Protons are used here since neutrons are unmeasured and isospin symmetry is expected at LHC energies.
Figure~\ref{fig-c5:coalescence} shows the measured coalescence parameters $B_2$ and $B_3$ as a function of the transverse momentum scaled by the mass number $A$ for Pb--Pb collisions  at $\mathbf{\sqrt{{\textit s}_{\rm NN}}} = 5.02$ TeV. An ordering of the coalescence parameters with collision centrality, from higher $B_A$ values in peripheral collisions to lower $B_A$ values in the central ones is clearly visible.
This trend with centrality is explained in the coalescence model framework as a consequence of the increasing radius $R$ of the source from peripheral to central events~\cite{Bellini:2018epz,Sun:2018mqq, Blum:2017qnn}.
Similarly, the decrease of $R$ with increasing momentum as measured with two-proton correlations~\cite{ALICE:2015hvw} can also explain the increase of the coalescence parameters with momentum observed in Fig.~\ref{fig-c5:coalescence}, as already seen in small collision systems~\cite{ALICE:2021mfm}.
Notably, the space-momentum correlations between nucleons restrict the volume that is effectively used for coalescence and makes it  smaller than the total volume. This means that particles at higher momentum probe a smaller region in the radially expanding system.

\begin{figure}[!t]
 	\begin{center}
 		\includegraphics[width=0.49\textwidth]{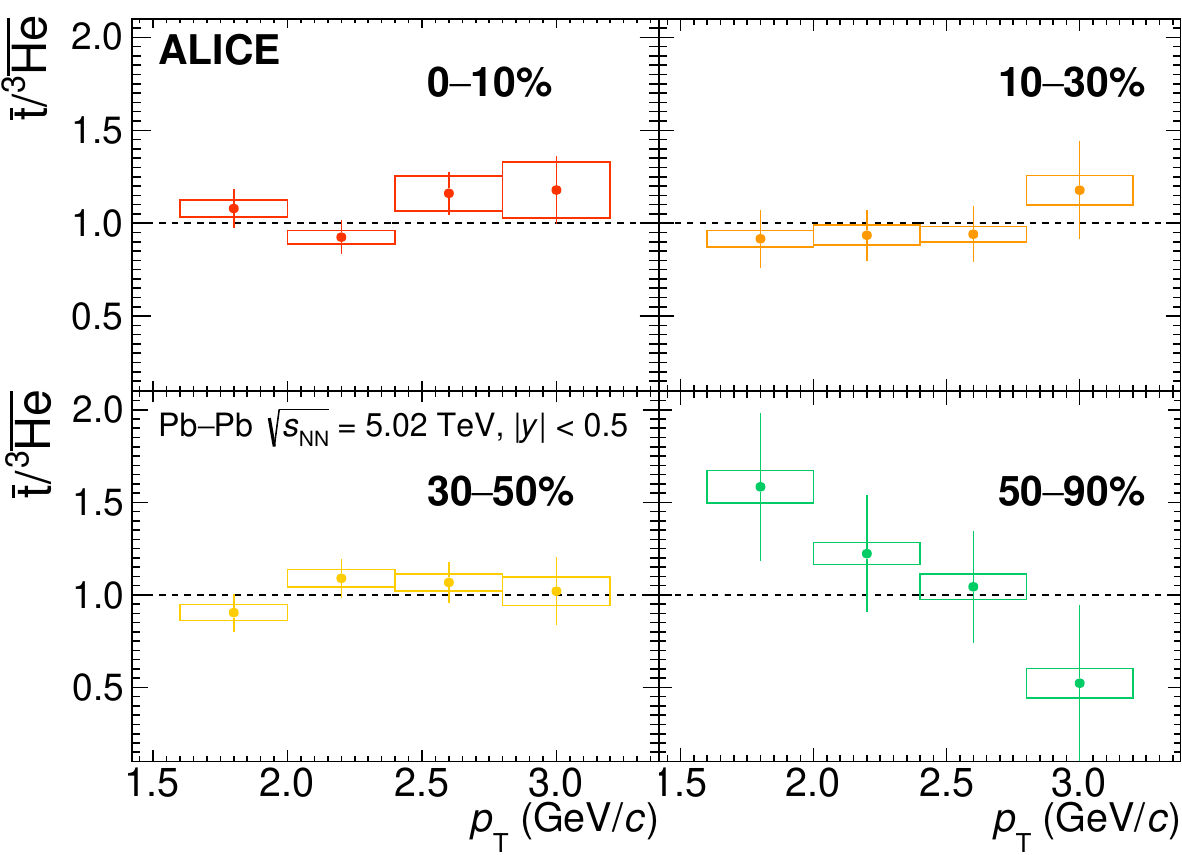}
 		\includegraphics[width=0.49\textwidth]{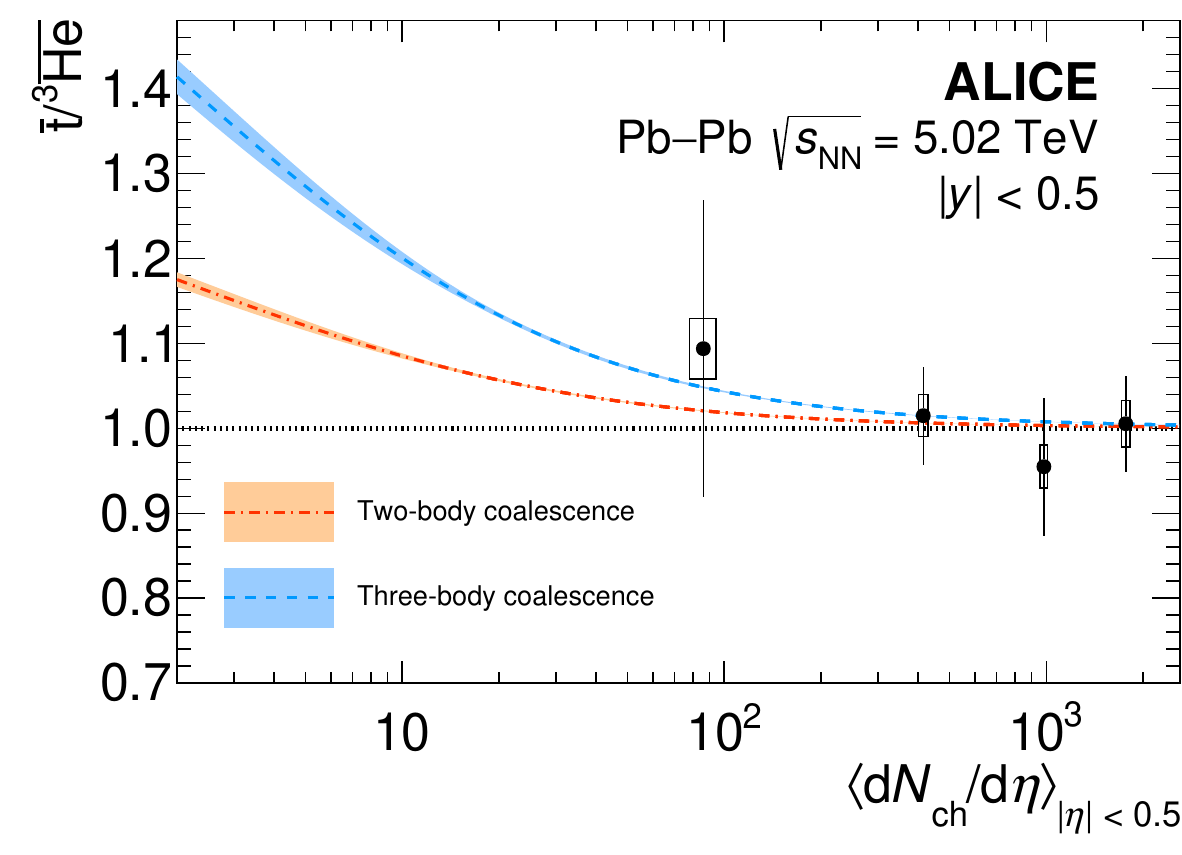}
 		\caption{\label{ttohe_ratios}Left: ratios of transverse momentum spectra of \antiHH and \antiHee in different centrality intervals. Right: multiplicity dependence of the average \antiHH/\antiHee ratio compared with the coalescence model expectations (two-body coalescence in orange and three-body coalescence in blue)~\cite{Sun:2018mqq}. The open boxes represent the total systematic uncertainties, while the vertical lines are the statistical ones.}
 	\end{center}
 \end{figure}

The ratio between the production yields of \antiHee and \antiHH provides another powerful test of the coalescence predictions~\cite{Sun:2018mqq}. This model gives two predictions for the formation, one assumes that the $A=3$ nuclei are formed from three nucleons (called three-body coalescence in the following) and the other assumes the formation of the nucleus from a deuteron and a nucleon (two-body coalescence). Figure~\ref{ttohe_ratios} shows on the left the measured ratios as a function of transverse momentum in different centrality classes. The ratios are flat with $p_{\mathrm{T}}$ within uncertainties. The average ratio \antiHH/\antiHee over the measured transverse momenta is shown in Fig.~\ref{ttohe_ratios} on the right, as a function of the charged-particle multiplicity $\langle {\rm d}N_{\rm ch}/{\rm d}\eta \rangle$ determined in a pseudorapidity range of $|\eta_{\rm lab}| < 0.5$ in the laboratory system.
While the SHM expectation for this ratio is very close to 1, the predictions from the coalescence model~\cite{Sun:2018mqq} deviate from unity due to the difference in the wave functions of the two nuclei. However, within the current statistical and systematic uncertainties, it is not possible to conclude which flavour of coalescence model is favoured by the measurement nor if there is any significant departure from the SHM expectations.

\begin{figure}[tp]
  \centering
 \includegraphics[width=\textwidth]{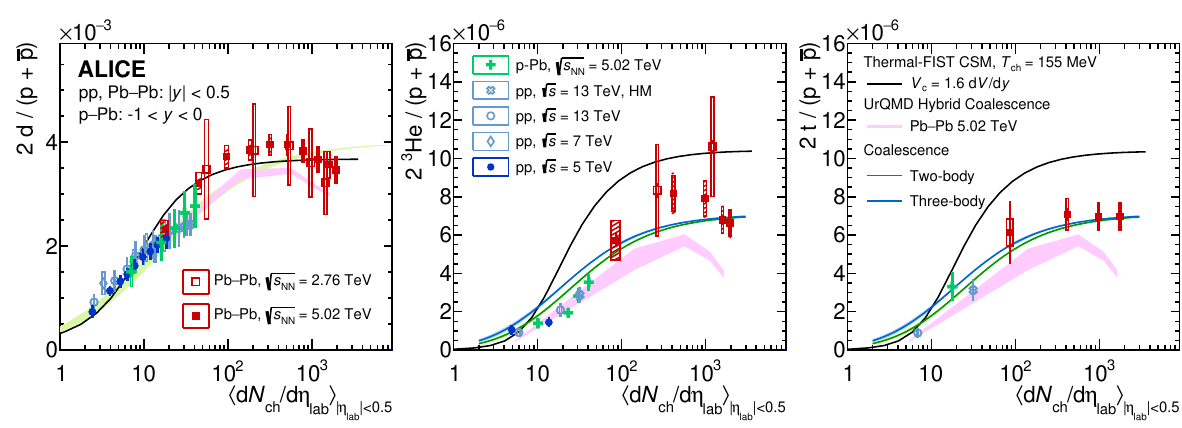}
  \caption{Integrated deuteron (left),  \Hee (middle), and triton yields (right) over  proton yields as a function of charged-particle multiplicity $\langle\textrm{d}N/\textrm{d}\eta_{\textrm{ch}}\rangle$ for pp, p--Pb, and Pb--Pb collisions measured by the ALICE Collaboration. The boxes represent the uncorrelated systematic uncertainties, while the vertical lines are the statistical ones. The shaded boxes represent the centrality-correlated uncertainties. In addition, the data are compared to the Thermal-FIST CSM (canonical statistical model) at 155\,MeV with a correlation volume of $V_\mathrm{c} = 1.6~{\rm d}V/{\rm d}y$ shown as a black line~\cite{Vovchenko:2018fiy}, the two coalescence approaches displayed in green (two-body coalescence) and in blue (three-body coalescence)~\cite{Sun:2018mqq} and UrQMD hybrid coalescence shown as a purple line~\cite{Reichert:2022mek}.}
  \label{fig:dpratio}
\end{figure}

Figure~\ref{fig:dpratio} shows the measured d/p, \Hee/p, and \HH/p for different collision systems at different energies.
Both the coalescence models (using the two different approaches, i.e. two-body and three-body coalescence~\cite{Sun:2018mqq}, and on top of UrQMD~\cite{Sombun:2018yqh,Reichert:2022mek}) and the Canonical SHM (CSM)~\cite{Vovchenko:2018fiy} capture qualitatively the observed trend with multiplicity. The UrQMD model uses a hybrid approach where nuclei are ultimately produced by coalescence. However, none of the model curves are able to explain quantitatively all the data points.
In all cases, a decrease in the ratios is observed from central Pb--Pb collisions toward peripheral Pb--Pb collisions. In particular, in the case of the d/p ratio, this depletion is significant when considering only the uncorrelated uncertainties. Such an effect is expected in transport codes modelling interactions of the nuclei in the rescattering phase following the hadron formation~\cite{Sombun:2018yqh}.

In the SHM assuming the grand canonical ensemble, the nucleus-to-proton ratios are fixed by the temperature of the source, thus they are expected to stay constant as a function of charged-particle multiplicity. However, when assuming a canonical ensemble and the exact conservation of baryon number over a defined volume, the nucleus-to-proton ratios increase from low to high multiplicities. The extension of the conservation volume was studied via event-by-event correlation measurement and it was found to be $V_\mathrm{c} = (1.6\pm0.2)~\dd V/\dd y$~\cite{ALICE:2022amd}. The studies presented here thus show that a small conservation volume is needed to describe deuteron-to-proton ratio in peripheral Pb--Pb collisions within the SHM, while the production of protons requires significantly larger values of around $V_\mathrm{c} =3\mbox{--}5~\dd V/\dd y$~\cite{ALICE:2019nbs}. At the moment, there is no configuration of the SHM that is able to describe simultaneously protons and nuclei with a single set of parameters in peripheral Pb--Pb and smaller collision systems. The decrease of the nucleus over proton ratios going towards smaller multiplicities is also expected in coalescence models, where it is caused by the evolution of the system size with multiplicity.

\section{Summary and conclusion}
\label{sec:summary}
In this article, we have presented comprehensive measurements of the production of (anti)nuclei in  Pb--Pb collisions at $\sqrt{s_{\textrm{NN}}} = 5.02$~TeV including the first antitriton measurement in Pb--Pb collisions at LHC energies. 
The obtained results follow the trends established at lower collision energy, but show a much larger constraining power on models thanks to significantly smaller systematic and statistical uncertainties.

For deuterons, the number of studied centrality intervals was largely increased compared to previous ALICE studies and demonstrate the strong increase of the radial flow when going from peripheral to central events. 
This is similarly visible for \Hee and \HH, but with a lower number of centrality intervals. The extracted spectra for \HH and \Hee agree well in the overlap region of both spectra. Only for the most peripheral interval a slight deviation from unity is visible, which is also expected by coalescence models, where the deviation at low multiplicities is expected due to the different spatial wave functions of \HH and \Hee~\cite{Bellini:2018epz,Sun:2018mqq}. This will be constrained better with high-statistics data from Run 3, using all available collision systems, i.e.\ pp, p--Pb, and Pb--Pb. 

The yield ratios of d/p as a function of charged-particle multiplicity agree well with both expectations, i.e. coalescence and thermal models. Notably, the deuteron-over-proton ratio requires a small correlation volume within the SHM with respect to net-proton fluctuation measurements. For \Hee the data lie at low multiplicity slightly closer to the coalescence expectations, and for high multiplicities corresponding to Pb--Pb the data lie between thermal and coalescence models. In contrast, for the triton the data points are much closer to the coalescence model with multiplicities in Pb--Pb collisions. Recently, several works appeared~\cite{Donigus:2022haq,Sharma:2022poi,Vovchenko:2022xil} that each try to improve the SHM in particular multiplicity regions. They give a good description in the region they are applied to, but they are not applicable in the full multiplicity range investigated here. 

The presented data, even though they are much more precise than previous results, still do not allow for a strong conclusion about the dominant production mechanism. More differential studies, in particular also those involving additional (hyper)nuclei, such as ${}^{4}$He and ${}^{3}_{\Lambda}$H will help to understand better the processes underlying the formation of composite objects.

The ongoing Run 3 of the LHC with the upgraded ALICE apparatus
will allow for such more precise studies of (anti)(hyper)nuclei production and for the extension to mass $A=4$ hypernuclei in Pb--Pb collisions~\cite{Dainese:2703572}.

\newenvironment{acknowledgement}{\relax}{\relax}
\begin{acknowledgement}
\section*{Acknowledgements}

The ALICE Collaboration would like to thank all its engineers and technicians for their invaluable contributions to the construction of the experiment and the CERN accelerator teams for the outstanding performance of the LHC complex.
The ALICE Collaboration gratefully acknowledges the resources and support provided by all Grid centres and the Worldwide LHC Computing Grid (WLCG) collaboration.
The ALICE Collaboration acknowledges the following funding agencies for their support in building and running the ALICE detector:
A. I. Alikhanyan National Science Laboratory (Yerevan Physics Institute) Foundation (ANSL), State Committee of Science and World Federation of Scientists (WFS), Armenia;
Austrian Academy of Sciences, Austrian Science Fund (FWF): [M 2467-N36] and Nationalstiftung f\"{u}r Forschung, Technologie und Entwicklung, Austria;
Ministry of Communications and High Technologies, National Nuclear Research Center, Azerbaijan;
Conselho Nacional de Desenvolvimento Cient\'{\i}fico e Tecnol\'{o}gico (CNPq), Financiadora de Estudos e Projetos (Finep), Funda\c{c}\~{a}o de Amparo \`{a} Pesquisa do Estado de S\~{a}o Paulo (FAPESP) and Universidade Federal do Rio Grande do Sul (UFRGS), Brazil;
Bulgarian Ministry of Education and Science, within the National Roadmap for Research Infrastructures 2020¿2027 (object CERN), Bulgaria;
Ministry of Education of China (MOEC) , Ministry of Science \& Technology of China (MSTC) and National Natural Science Foundation of China (NSFC), China;
Ministry of Science and Education and Croatian Science Foundation, Croatia;
Centro de Aplicaciones Tecnol\'{o}gicas y Desarrollo Nuclear (CEADEN), Cubaenerg\'{\i}a, Cuba;
Ministry of Education, Youth and Sports of the Czech Republic, Czech Republic;
The Danish Council for Independent Research | Natural Sciences, the VILLUM FONDEN and Danish National Research Foundation (DNRF), Denmark;
Helsinki Institute of Physics (HIP), Finland;
Commissariat \`{a} l'Energie Atomique (CEA) and Institut National de Physique Nucl\'{e}aire et de Physique des Particules (IN2P3) and Centre National de la Recherche Scientifique (CNRS), France;
Bundesministerium f\"{u}r Bildung und Forschung (BMBF) and GSI Helmholtzzentrum f\"{u}r Schwerionenforschung GmbH, Germany;
General Secretariat for Research and Technology, Ministry of Education, Research and Religions, Greece;
National Research, Development and Innovation Office, Hungary;
Department of Atomic Energy Government of India (DAE), Department of Science and Technology, Government of India (DST), University Grants Commission, Government of India (UGC) and Council of Scientific and Industrial Research (CSIR), India;
National Research and Innovation Agency - BRIN, Indonesia;
Istituto Nazionale di Fisica Nucleare (INFN), Italy;
Japanese Ministry of Education, Culture, Sports, Science and Technology (MEXT) and Japan Society for the Promotion of Science (JSPS) KAKENHI, Japan;
Consejo Nacional de Ciencia (CONACYT) y Tecnolog\'{i}a, through Fondo de Cooperaci\'{o}n Internacional en Ciencia y Tecnolog\'{i}a (FONCICYT) and Direcci\'{o}n General de Asuntos del Personal Academico (DGAPA), Mexico;
Nederlandse Organisatie voor Wetenschappelijk Onderzoek (NWO), Netherlands;
The Research Council of Norway, Norway;
Commission on Science and Technology for Sustainable Development in the South (COMSATS), Pakistan;
Pontificia Universidad Cat\'{o}lica del Per\'{u}, Peru;
Ministry of Education and Science, National Science Centre and WUT ID-UB, Poland;
Korea Institute of Science and Technology Information and National Research Foundation of Korea (NRF), Republic of Korea;
Ministry of Education and Scientific Research, Institute of Atomic Physics, Ministry of Research and Innovation and Institute of Atomic Physics and University Politehnica of Bucharest, Romania;
Ministry of Education, Science, Research and Sport of the Slovak Republic, Slovakia;
National Research Foundation of South Africa, South Africa;
Swedish Research Council (VR) and Knut \& Alice Wallenberg Foundation (KAW), Sweden;
European Organization for Nuclear Research, Switzerland;
Suranaree University of Technology (SUT), National Science and Technology Development Agency (NSTDA), Thailand Science Research and Innovation (TSRI) and National Science, Research and Innovation Fund (NSRF), Thailand;
Turkish Energy, Nuclear and Mineral Research Agency (TENMAK), Turkey;
National Academy of  Sciences of Ukraine, Ukraine;
Science and Technology Facilities Council (STFC), United Kingdom;
National Science Foundation of the United States of America (NSF) and United States Department of Energy, Office of Nuclear Physics (DOE NP), United States of America.
In addition, individual groups or members have received support from:
Marie Sk\l{}odowska Curie, European Research Council, Strong 2020 - Horizon 2020 (grant nos. 950692, 824093, 896850), European Union;
Academy of Finland (Center of Excellence in Quark Matter) (grant nos. 346327, 346328), Finland;
Programa de Apoyos para la Superaci\'{o}n del Personal Acad\'{e}mico, UNAM, Mexico.
\end{acknowledgement}

\bibliographystyle{utphys}   
\bibliography{bibliography}

\newpage

\appendix
\section{The ALICE Collaboration}
\label{app:collab}
\begin{flushleft} 
\small

S.~Acharya\,\orcidlink{0000-0002-9213-5329}\,$^{\rm 125}$, 
D.~Adamov\'{a}\,\orcidlink{0000-0002-0504-7428}\,$^{\rm 86}$, 
A.~Adler$^{\rm 69}$, 
G.~Aglieri Rinella\,\orcidlink{0000-0002-9611-3696}\,$^{\rm 32}$, 
M.~Agnello\,\orcidlink{0000-0002-0760-5075}\,$^{\rm 29}$, 
N.~Agrawal\,\orcidlink{0000-0003-0348-9836}\,$^{\rm 50}$, 
Z.~Ahammed\,\orcidlink{0000-0001-5241-7412}\,$^{\rm 132}$, 
S.~Ahmad\,\orcidlink{0000-0003-0497-5705}\,$^{\rm 15}$, 
S.U.~Ahn\,\orcidlink{0000-0001-8847-489X}\,$^{\rm 70}$, 
I.~Ahuja\,\orcidlink{0000-0002-4417-1392}\,$^{\rm 37}$
% ...

\section*{Affiliation Notes}

$^{\rm I}$ Deceased\\
$^{\rm II}$ Also at: Max-Planck-Institut f\"{u}r Physik, Munich, Germany\\
$^{\rm III}$ Also at: Italian National Agency for New Technologies, Energy and Sustainable Economic Development (ENEA), Bologna, Italy\\
$^{\rm IV}$ Also at: Dipartimento DET del Politecnico di Torino, Turin, Italy\\
$^{\rm V}$ Also at: Department of Applied Physics, Aligarh Muslim University, Aligarh, India\\
$^{\rm VI}$ Also at: Institute of Theoretical Physics, University of Wroclaw, Poland\\
$^{\rm VII}$ Also at: An institution covered by a cooperation agreement with CERN\\
$^{\rm VIII}$ Also at: Indian Institute of Science Education and Research (IISER) Berhampur, Odisha, India\\

\section*{Collaboration Institutes}

$^{1}$ A.I. Alikhanyan National Science Laboratory (Yerevan Physics Institute) Foundation, Yerevan, Armenia\\
$^{2}$ AGH University of Science and Technology, Cracow, Poland\\
$^{3}$ Bogolyubov Institute for Theoretical Physics, National Academy of Sciences of Ukraine, Kiev, Ukraine\\
$^{4}$ Bose Institute, Department of Physics  and Centre for Astroparticle Physics and Space Science (CAPSS), Kolkata, India\\
$^{5}$ California Polytechnic State University, San Luis Obispo, California, United States\\
$^{6}$ Central China Normal University, Wuhan, China\\
$^{7}$ Centro de Aplicaciones Tecnol\'{o}gicas y Desarrollo Nuclear (CEADEN), Havana, Cuba\\
$^{8}$ Centro de Investigaci\'{o}n y de Estudios Avanzados (CINVESTAV), Mexico City and M\'{e}rida, Mexico\\
$^{9}$ Chicago State University, Chicago, Illinois, United States\\
$^{10}$ China Institute of Atomic Energy, Beijing, China\\
$^{11}$ Chungbuk National University, Cheongju, Republic of Korea\\
$^{12}$ Comenius University Bratislava, Faculty of Mathematics, Physics and Informatics, Bratislava, Slovak Republic\\
$^{13}$ COMSATS University Islamabad, Islamabad, Pakistan\\
$^{14}$ Creighton University, Omaha, Nebraska, United States\\
$^{15}$ Department of Physics, Aligarh Muslim University, Aligarh, India\\
$^{16}$ Department of Physics, Pusan National University, Pusan, Republic of Korea\\
$^{17}$ Department of Physics, Sejong University, Seoul, Republic of Korea\\
$^{18}$ Department of Physics, University of California, Berkeley, California, United States\\
$^{19}$ Department of Physics, University of Oslo, Oslo, Norway\\
$^{20}$ Department of Physics and Technology, University of Bergen, Bergen, Norway\\
$^{21}$ Dipartimento di Fisica, Universit\`{a} di Pavia, Pavia, Italy\\
$^{22}$ Dipartimento di Fisica dell'Universit\`{a} and Sezione INFN, Cagliari, Italy\\
$^{23}$ Dipartimento di Fisica dell'Universit\`{a} and Sezione INFN, Trieste, Italy\\
$^{24}$ Dipartimento di Fisica dell'Universit\`{a} and Sezione INFN, Turin, Italy\\
$^{25}$ Dipartimento di Fisica e Astronomia dell'Universit\`{a} and Sezione INFN, Bologna, Italy\\
$^{26}$ Dipartimento di Fisica e Astronomia dell'Universit\`{a} and Sezione INFN, Catania, Italy\\
$^{27}$ Dipartimento di Fisica e Astronomia dell'Universit\`{a} and Sezione INFN, Padova, Italy\\
$^{28}$ Dipartimento di Fisica `E.R.~Caianiello' dell'Universit\`{a} and Gruppo Collegato INFN, Salerno, Italy\\
$^{29}$ Dipartimento DISAT del Politecnico and Sezione INFN, Turin, Italy\\
$^{30}$ Dipartimento di Scienze MIFT, Universit\`{a} di Messina, Messina, Italy\\
$^{31}$ Dipartimento Interateneo di Fisica `M.~Merlin' and Sezione INFN, Bari, Italy\\
$^{32}$ European Organization for Nuclear Research (CERN), Geneva, Switzerland\\
$^{33}$ Faculty of Electrical Engineering, Mechanical Engineering and Naval Architecture, University of Split, Split, Croatia\\
$^{34}$ Faculty of Engineering and Science, Western Norway University of Applied Sciences, Bergen, Norway\\
$^{35}$ Faculty of Nuclear Sciences and Physical Engineering, Czech Technical University in Prague, Prague, Czech Republic\\
$^{36}$ Faculty of Physics, Sofia University, Sofia, Bulgaria\\
$^{37}$ Faculty of Science, P.J.~\v{S}af\'{a}rik University, Ko\v{s}ice, Slovak Republic\\
$^{38}$ Frankfurt Institute for Advanced Studies, Johann Wolfgang Goethe-Universit\"{a}t Frankfurt, Frankfurt, Germany\\
$^{39}$ Fudan University, Shanghai, China\\
$^{40}$ Gangneung-Wonju National University, Gangneung, Republic of Korea\\
$^{41}$ Gauhati University, Department of Physics, Guwahati, India\\
$^{42}$ Helmholtz-Institut f\"{u}r Strahlen- und Kernphysik, Rheinische Friedrich-Wilhelms-Universit\"{a}t Bonn, Bonn, Germany\\
$^{43}$ Helsinki Institute of Physics (HIP), Helsinki, Finland\\
$^{44}$ High Energy Physics Group,  Universidad Aut\'{o}noma de Puebla, Puebla, Mexico\\
$^{45}$ Horia Hulubei National Institute of Physics and Nuclear Engineering, Bucharest, Romania\\
$^{46}$ Indian Institute of Technology Bombay (IIT), Mumbai, India\\
$^{47}$ Indian Institute of Technology Indore, Indore, India\\
$^{48}$ INFN, Laboratori Nazionali di Frascati, Frascati, Italy\\
$^{49}$ INFN, Sezione di Bari, Bari, Italy\\
$^{50}$ INFN, Sezione di Bologna, Bologna, Italy\\
$^{51}$ INFN, Sezione di Cagliari, Cagliari, Italy\\
$^{52}$ INFN, Sezione di Catania, Catania, Italy\\
$^{53}$ INFN, Sezione di Padova, Padova, Italy\\
$^{54}$ INFN, Sezione di Pavia, Pavia, Italy\\
$^{55}$ INFN, Sezione di Torino, Turin, Italy\\
$^{56}$ INFN, Sezione di Trieste, Trieste, Italy\\
$^{57}$ Inha University, Incheon, Republic of Korea\\
$^{58}$ Institute for Gravitational and Subatomic Physics (GRASP), Utrecht University/Nikhef, Utrecht, Netherlands\\
$^{59}$ Institute of Experimental Physics, Slovak Academy of Sciences, Ko\v{s}ice, Slovak Republic\\
$^{60}$ Institute of Physics, Homi Bhabha National Institute, Bhubaneswar, India\\
$^{61}$ Institute of Physics of the Czech Academy of Sciences, Prague, Czech Republic\\
$^{62}$ Institute of Space Science (ISS), Bucharest, Romania\\
$^{63}$ Institut f\"{u}r Kernphysik, Johann Wolfgang Goethe-Universit\"{a}t Frankfurt, Frankfurt, Germany\\
$^{64}$ Instituto de Ciencias Nucleares, Universidad Nacional Aut\'{o}noma de M\'{e}xico, Mexico City, Mexico\\
$^{65}$ Instituto de F\'{i}sica, Universidade Federal do Rio Grande do Sul (UFRGS), Porto Alegre, Brazil\\
$^{66}$ Instituto de F\'{\i}sica, Universidad Nacional Aut\'{o}noma de M\'{e}xico, Mexico City, Mexico\\
$^{67}$ iThemba LABS, National Research Foundation, Somerset West, South Africa\\
$^{68}$ Jeonbuk National University, Jeonju, Republic of Korea\\
$^{69}$ Johann-Wolfgang-Goethe Universit\"{a}t Frankfurt Institut f\"{u}r Informatik, Fachbereich Informatik und Mathematik, Frankfurt, Germany\\
$^{70}$ Korea Institute of Science and Technology Information, Daejeon, Republic of Korea\\
$^{71}$ KTO Karatay University, Konya, Turkey\\
$^{72}$ Laboratoire de Physique des 2 Infinis, Ir\`{e}ne Joliot-Curie, Orsay, France\\
$^{73}$ Laboratoire de Physique Subatomique et de Cosmologie, Universit\'{e} Grenoble-Alpes, CNRS-IN2P3, Grenoble, France\\
$^{74}$ Lawrence Berkeley National Laboratory, Berkeley, California, United States\\
$^{75}$ Lund University Department of Physics, Division of Particle Physics, Lund, Sweden\\
$^{76}$ Nagasaki Institute of Applied Science, Nagasaki, Japan\\
$^{77}$ Nara Women{'}s University (NWU), Nara, Japan\\
$^{78}$ National and Kapodistrian University of Athens, School of Science, Department of Physics , Athens, Greece\\
$^{79}$ National Centre for Nuclear Research, Warsaw, Poland\\
$^{80}$ National Institute of Science Education and Research, Homi Bhabha National Institute, Jatni, India\\
$^{81}$ National Nuclear Research Center, Baku, Azerbaijan\\
$^{82}$ National Research and Innovation Agency - BRIN, Jakarta, Indonesia\\
$^{83}$ Niels Bohr Institute, University of Copenhagen, Copenhagen, Denmark\\
$^{84}$ Nikhef, National institute for subatomic physics, Amsterdam, Netherlands\\
$^{85}$ Nuclear Physics Group, STFC Daresbury Laboratory, Daresbury, United Kingdom\\
$^{86}$ Nuclear Physics Institute of the Czech Academy of Sciences, Husinec-\v{R}e\v{z}, Czech Republic\\
$^{87}$ Oak Ridge National Laboratory, Oak Ridge, Tennessee, United States\\
$^{88}$ Ohio State University, Columbus, Ohio, United States\\
$^{89}$ Physics department, Faculty of science, University of Zagreb, Zagreb, Croatia\\
$^{90}$ Physics Department, Panjab University, Chandigarh, India\\
$^{91}$ Physics Department, University of Jammu, Jammu, India\\
$^{92}$ Physics Program and International Institute for Sustainability with Knotted Chiral Meta Matter (SKCM2), Hiroshima University, Hiroshima, Japan\\
$^{93}$ Physikalisches Institut, Eberhard-Karls-Universit\"{a}t T\"{u}bingen, T\"{u}bingen, Germany\\
$^{94}$ Physikalisches Institut, Ruprecht-Karls-Universit\"{a}t Heidelberg, Heidelberg, Germany\\
$^{95}$ Physik Department, Technische Universit\"{a}t M\"{u}nchen, Munich, Germany\\
$^{96}$ Politecnico di Bari and Sezione INFN, Bari, Italy\\
$^{97}$ Research Division and ExtreMe Matter Institute EMMI, GSI Helmholtzzentrum f\"ur Schwerionenforschung GmbH, Darmstadt, Germany\\
$^{98}$ Saga University, Saga, Japan\\
$^{99}$ Saha Institute of Nuclear Physics, Homi Bhabha National Institute, Kolkata, India\\
$^{100}$ School of Physics and Astronomy, University of Birmingham, Birmingham, United Kingdom\\
$^{101}$ Secci\'{o}n F\'{\i}sica, Departamento de Ciencias, Pontificia Universidad Cat\'{o}lica del Per\'{u}, Lima, Peru\\
$^{102}$ Stefan Meyer Institut f\"{u}r Subatomare Physik (SMI), Vienna, Austria\\
$^{103}$ SUBATECH, IMT Atlantique, Nantes Universit\'{e}, CNRS-IN2P3, Nantes, France\\
$^{104}$ Sungkyunkwan University, Suwon City, Republic of Korea\\
$^{105}$ Suranaree University of Technology, Nakhon Ratchasima, Thailand\\
$^{106}$ Technical University of Ko\v{s}ice, Ko\v{s}ice, Slovak Republic\\
$^{107}$ The Henryk Niewodniczanski Institute of Nuclear Physics, Polish Academy of Sciences, Cracow, Poland\\
$^{108}$ The University of Texas at Austin, Austin, Texas, United States\\
$^{109}$ Universidad Aut\'{o}noma de Sinaloa, Culiac\'{a}n, Mexico\\
$^{110}$ Universidade de S\~{a}o Paulo (USP), S\~{a}o Paulo, Brazil\\
$^{111}$ Universidade Estadual de Campinas (UNICAMP), Campinas, Brazil\\
$^{112}$ Universidade Federal do ABC, Santo Andre, Brazil\\
$^{113}$ University of Cape Town, Cape Town, South Africa\\
$^{114}$ University of Houston, Houston, Texas, United States\\
$^{115}$ University of Jyv\"{a}skyl\"{a}, Jyv\"{a}skyl\"{a}, Finland\\
$^{116}$ University of Kansas, Lawrence, Kansas, United States\\
$^{117}$ University of Liverpool, Liverpool, United Kingdom\\
$^{118}$ University of Science and Technology of China, Hefei, China\\
$^{119}$ University of South-Eastern Norway, Kongsberg, Norway\\
$^{120}$ University of Tennessee, Knoxville, Tennessee, United States\\
$^{121}$ University of the Witwatersrand, Johannesburg, South Africa\\
$^{122}$ University of Tokyo, Tokyo, Japan\\
$^{123}$ University of Tsukuba, Tsukuba, Japan\\
$^{124}$ University Politehnica of Bucharest, Bucharest, Romania\\
$^{125}$ Universit\'{e} Clermont Auvergne, CNRS/IN2P3, LPC, Clermont-Ferrand, France\\
$^{126}$ Universit\'{e} de Lyon, CNRS/IN2P3, Institut de Physique des 2 Infinis de Lyon, Lyon, France\\
$^{127}$ Universit\'{e} de Strasbourg, CNRS, IPHC UMR 7178, F-67000 Strasbourg, France, Strasbourg, France\\
$^{128}$ Universit\'{e} Paris-Saclay Centre d'Etudes de Saclay (CEA), IRFU, D\'{e}partment de Physique Nucl\'{e}aire (DPhN), Saclay, France\\
$^{129}$ Universit\`{a} degli Studi di Foggia, Foggia, Italy\\
$^{130}$ Universit\`{a} del Piemonte Orientale, Vercelli, Italy\\
$^{131}$ Universit\`{a} di Brescia, Brescia, Italy\\
$^{132}$ Variable Energy Cyclotron Centre, Homi Bhabha National Institute, Kolkata, India\\
$^{133}$ Warsaw University of Technology, Warsaw, Poland\\
$^{134}$ Wayne State University, Detroit, Michigan, United States\\
$^{135}$ Westf\"{a}lische Wilhelms-Universit\"{a}t M\"{u}nster, Institut f\"{u}r Kernphysik, M\"{u}nster, Germany\\
$^{136}$ Wigner Research Centre for Physics, Budapest, Hungary\\
$^{137}$ Yale University, New Haven, Connecticut, United States\\
$^{138}$ Yonsei University, Seoul, Republic of Korea\\
$^{139}$  Zentrum  f\"{u}r Technologie und Transfer (ZTT), Worms, Germany\\
$^{140}$ Affiliated with an institute covered by a cooperation agreement with CERN\\
$^{141}$ Affiliated with an international laboratory covered by a cooperation agreement with CERN.\\

\end{flushleft} 
\end{document}